\documentclass[pra,twoside,twocolumn,showpacs]{revtex4-1}
\usepackage{amsmath}
\usepackage{amssymb}
\usepackage{amsfonts}

\renewcommand{\vec}{\textbf}
\newcommand{\ket}[1]{|#1\rangle}
\newcommand{\bra}[1]{\langle#1|}
\newcommand{\bracket}[2]{\langle#1|#2\rangle}

\usepackage{graphicx}
\usepackage[amssymb]{SIunits}
\usepackage[active]{srcltx}

\begin{document}
\bibliographystyle{apsrev4-1}
\title{Relativistic Einstein-Podolsky-Rosen correlations and localization}

\author{Pawe{\l}{} Caban}
\email{P.Caban@merlin.phys.uni.lodz.pl}
\author{Jakub Rembieli{\'n}ski}
\email{jaremb@uni.lodz.pl}
\author{Patrycja Rybka}
\email{p.rybka@onet.pl}
\author{Piotr Witas}
\email{pjwitas@gmail.com}

\affiliation{Department of Theoretical Physics, University of Lodz\\
Pomorska 149/153, 90-236 {\L}{\'o}d{\'z}, Poland}
\date{\today}

\begin{abstract}
We calculate correlation functions for a relativistic
Einstein-Podolsky-Rosen-type experiment with massive Dirac
particles. We take the influence of the Newton-Wigner localization
into account and perform the calculations for a couple of physically
interesting states. 
\end{abstract}
\maketitle


Relativistic Einstein-Podolsky-Rosen-type (EPR-type) correlations have
attracted a lot of attention in recent years (see, e.g.,
Refs.~\cite{CR2006,FBHH2010,Czachor2010,SV2011,DV2012} 
and references therein). Such
correlations are especially interesting in systems of massive
particles. In this case relativistic spin correlation function can be
a nonmonotonic function of particle momenta \cite{CRW2009} (in
contrast to the case of photons, where polarization correlations
behaves like spin correlations in nonrelativistic quantum mechanics
\cite{Caban2007_photons}). However, previous discussions did not take
into account the localization of EPR particles inside detectors. Such
a localization usually accompanies the spin projection measurement.

The main purpose of our paper is to include the localization in the
discussion of EPR-type spin correlations. However, there are some
issues we have to overcome. First of all, various spin operators have
been used in the discussion of relativistic correlations (see, e.g.,
Refs.~\cite{Czachor1997_1,Terno2003,CR2005,SV2012,PTW2013_spin_WKB}). 
Secondly, the notion of localization is not
well-defined in relativistic quantum mechanics \cite{cab_Bacry1988}.
These two
problems are connected because spin can be defined as a difference
between total and orbital angular momentum
[Eq.~(\ref{spin_difference})], where orbital angular momentum is defined
with the help of position operator, see also \cite{CRW2009}.

In our recent papers
\cite{CRW_2013_Dirac_spin_PRA} we have shown that
the most apriopriate spin operator is an operator connected with the
Newton-Wigner (NW) position operator \cite{cab_NW1949}. On the other hand,
NW position operator, although noncovariant, seems to be
the best proposition for a relativistic position operator. Therefore,
in this paper we use the NW localization and the
corresponding spin operator. We consider EPR correlations in a scalar
state of two spin-1/2 particles assuming that spin projection
measurements take place in finite-volume regions (detectors). 
We derive a general formula for the correlation function and then
consider some special cases.


The carrier space $\mathcal{H}$
of the unitary representation of the Poincar\'e group for spin 1/2 is
spanned by the eigenvectors of four-momentum $\hat{P}^{\mu}$.
These states are denoted as $\ket{p,\sigma}$, $\sigma=\pm1/2$ and are
normalized as follows: 
\begin{equation}
 \label{eq:norma}
 \bracket{p,\sigma}{k,\lambda} = 2 p^{0} \delta^{3}(\vec{p}-\vec{k})
 \delta_{\sigma\lambda}.
\end{equation}
Under the action of the Lorentz group the basis states transforms as
\begin{equation}
 U(\Lambda) \ket{p,\sigma} = 
 \mathcal{D}^{1/2}_{\lambda\sigma}(R(\Lambda,p)) 
 \ket{\Lambda p,\lambda}, 
\end{equation}
where $\mathcal{D}^{1/2}$ is the spin-1/2 representation of the 
rotation group and $R(\Lambda,p)$ is a Wigner rotation. 

Lorentz-covariant singlet two-particle state has the form (see
e.g.~\cite{CR2006})  
\begin{equation}
 \label{eg:stan_1}
 \ket{\varphi} = 
 \int \frac{d^{3}\vec{k}}{2k^{0}} \frac{d^{3}\vec{p}}{2p^{0}}
 \varphi(k,p) \mathcal{M}_{\sigma\lambda}(k,p) 
 \ket{k,\sigma}\otimes\ket{p,\lambda},
\end{equation}
where $\varphi(k,p)$ is a scalar function and for the particles with
the same mass $m$ the matrix $\mathcal{M}$ reads:
\begin{multline}
 \label{amp}
 \mathcal{M}(k,p)_{\sigma\lambda} =
 -i \Big( 2m\sqrt{(m+p^{0})(m+k^{0})} \Big)^{-1}
 \\ \times
 \Big( \big[(m+k^{0})(m+p^{0}) - \vec{k}\cdot\vec{p} 
 - i\boldsymbol{\sigma}\cdot(\vec{k}\times\vec{p})\big]\sigma_{2}
  \Big)_{\sigma\lambda}.
\end{multline}
In the above equation
$\boldsymbol{\sigma}=(\sigma_1,\sigma_2,\sigma_3)$ and $\sigma_i$ are 
the standard Pauli matrices. For the details on the definition and
properties of Lorentz-covariant states we refer the reader to
\cite{CR2006}.


It is well-known that the spin square operator can be uniquely defined
in terms of the generators of the Poincar\'e group as
\begin{equation}
 \hat{\vec{S}}^2=-\frac{1}{m^2}\hat{W}^{\mu}\hat{W}_{\mu},
\end{equation}
where $\hat{W}^{\mu}$ is the Pauli-Lubanski four-vector:
$\hat{W}^{\mu}=
\frac{1}{2}\epsilon^{\nu\alpha\beta\mu}\hat{P}_{\nu}\hat{J}_{\alpha\beta}$,
and $\hat{J}_{\alpha\beta}$ are the generators of the Lorentz
group.
In spite of that the definition of the relativistic spin operator has
been widely discussed in the literature (see, e.g., 
\cite{FW1950,cab_BLT1969,Czachor1997_1,Terno2003,%
CRW_2013_Dirac_spin_PRA} and references therein).
In the enveloping algebra of the Lie algebra of the Poincar\'e group
we can define a relativistic spin operator
\begin{equation}
 \label{eq:definicja_operatora_spinu}
 \Hat{\vec{S}} =
 \frac{1}{m}\left(\hat{\vec{W}}
 -\hat{W}^0\frac{\hat{\vec{P}}}{\hat{P}^0+m}\right).
\end{equation}
This operator is linear in the components of $\Hat{W}^\mu$,
transforms like a vector under
rotations and like a pseudo-vector under reflections, commutes with
space-time observables and fulfills the standard canonical commutation
relations (for the details see, e.g., \cite{cab_BLT1969}). 
Moreover, it has
been shown \cite{CRW_2013_Dirac_spin} that the spin operator
(\ref{eq:definicja_operatora_spinu}) 
is equivalent to the action of the mean-spin operator
introduced by Foldy and Wouthuysen \cite{FW1950} in the Dirac theory.

The action of the operator (\ref{eq:definicja_operatora_spinu}) on the
basis vectors is the following: 
\begin{equation}
 \hat{\vec{S}} \ket{p,\sigma} =
 \frac{\boldsymbol{\sigma}_{\sigma^\prime\sigma}}{2}
 \ket{p,\sigma^\prime}.
 \label{spin}
\end{equation}

On the other hand, spin can be defined as a difference between total
and orbital angular momentum
\begin{equation}
 \hat{\vec{S}} = \hat{\vec{J}} - \hat{\vec{X}}\times\hat{\vec{P}},
 \label{spin_difference}
\end{equation}
where $\Hat{J}^i = \epsilon^{ijk} \Hat{J}^{jk}$ and $\Hat{X}$ is a
position operator. However, there does not exist unambigously defined
position operator in relativistic quantum mechanics
\cite{cab_Bacry1988}. The best 
proposition seems to be the Newton-Wigner position operator
\cite{cab_NW1949}. It appears that when we insert the Newton-Wigner
position operator as $\Hat{X}$ in Eq.~(\ref{spin_difference}) than the
resulting spin operator coincides with the spin defined in
Eq.~(\ref{eq:definicja_operatora_spinu}). 

For further convenience we briefly remind some properties of the
Newton-Wigner position operator.
Arbitrary one-particle state can be written as
\begin{equation}
 \ket{\psi} = \int \frac{d^3 \vec{p}}{2p^0}
 \psi_\sigma(p) \ket{p,\sigma}.
\end{equation}
The action of the Newton-Wigner position operator on
wave function in the momentum representation has the well-known form 
\begin{equation}
 \hat{\vec{X}} \psi_{\sigma}(p) = 
 \Big( i\boldsymbol{\nabla}_{\vec{p}} 
 - \frac{1}{2} \frac{i\vec{p}}{\vec{p}^2+m^2}\Big)\psi_{\sigma}(p).
\end{equation}
The eigenstates of this operator are
\begin{equation}
 \label{eq:eigenstate}
 \ket{\vec{x},\sigma} = 
 (2\pi)^{-3/2} \int \frac{d^3\vec{p}}{2p^0} \sqrt{2p^0}
 e^{-i\vec{p}\cdot\vec{x}} \ket{p,\sigma}.
\end{equation}
Consequently, we introduce a projector on a region $\Omega$
\begin{equation}
 \label{projector}
 \hat{\Pi}_\Omega =
 \sum_{\sigma} \int_{\Omega} d^3 \vec{x}\,
 \ket{\vec{x},\sigma}\bra{\vec{x},\sigma}.
\end{equation}
Using Eq.~(\ref{eq:eigenstate}) we get
\begin{multline}
 \label{eq:pi_omega}
 \hat{\Pi}_\Omega =
 \int 
 \frac{d^3\vec{p}^\prime}{\sqrt{2p^{\prime0}}}
 \frac{d^3\vec{p}}{\sqrt{2p^0}} 
 \Delta_{\Omega}(\vec{p}^\prime-\vec{p})
 \sum_{\sigma} \ket{p^\prime,\sigma} \bra{p,\sigma},
\end{multline}
where
\begin{equation}
 \label{eq:Delta_p_minus_k}
 \Delta_{\Omega}(\vec{p}^\prime-\vec{p}) =
 \frac{1}{(2\pi)^3} \int_{\Omega} d^3\vec{x}\,
 e^{-i(\vec{p}^\prime-\vec{p})\cdot\vec{x}}.
\end{equation}
Notice that
\begin{equation}
 \Delta_{\mathbb{R}^3}(\vec{p}^\prime-\vec{p}) =
 \delta^{3}(\vec{p}^\prime-\vec{p}).
\end{equation}
Spin projection measurement in the direction $\vec{n}$ in the
region $\Omega$ is described by the following observable:
\begin{equation}
 \vec{n}\cdot\hat{\vec{S}}_{\Omega} =
 \big(\vec{n}\cdot\hat{\vec{S}}\big) \hat{\Pi}_\Omega,
 \label{S_Omega}
\end{equation}
where $\hat{\Pi}_\Omega$ and $\hat{\vec{S}}$ are given by
Eqs.~(\ref{projector}) and (\ref{spin}), respectively.


Now let us consider an EPR-type experiment. That is, we assume that
two particles are produced in the state (\ref{eg:stan_1}) and sent to
two distant observers, Alice and Bob. Alice (Bob) measures the spin
projection in the direction $\vec{a}$ ($\vec{b}$) provided that her
(his) particle is inside the region A (B). It means that Alice
measures the observable
$\big(\vec{a}\cdot\hat{\vec{S}}\big) \hat{\Pi}_A$ while Bob
$\big(\vec{b}\cdot\hat{\vec{S}}\big) \hat{\Pi}_B$.
The normalized correlation function reads:
\begin{equation}
 \label{eq:korelacje_omega}
 \mathcal{C}_{\varphi}^{AB}(\vec{a},\vec{b}) = 4
 \frac{\bra{\varphi}\hat{\Pi}_A \big(\vec{a}\cdot\hat{\vec{S}}\big)
 \otimes \big(\vec{b}\cdot\hat{\vec{S}}\big) \hat{\Pi}_B
 \ket{\varphi}}{\bra{\varphi} \hat{\Pi}_A\otimes\hat{\Pi}_B
 \ket{\varphi}}.
\end{equation}
The form of the denominator in Eq.~(\ref{eq:korelacje_omega})
corresponds to the fact that we take into account only the pairs that
are actually found inside the detectors and the function has to be
appropriately normalized.
Using Eqs.~(\ref{eg:stan_1},\ref{spin},\ref{eq:pi_omega}) and
(\ref{S_Omega}) we get 
\begin{multline}
 \label{eq:korelacja_licznik}
 \bra{\varphi} \hat{\Pi}_A \big(\vec{a}\cdot\hat{\vec{S}}\big)
 \otimes \big(\vec{b}\cdot\hat{\vec{S}}\big) 
 \hat{\Pi}_B \ket{\varphi} 
 \\ =
 \frac{1}{4}
 \int \frac{d^3\vec{k}^\prime\,d^3\vec{k}\,
 d^3\vec{p}^\prime\,d^3\vec{p}}{\sqrt{2k^{\prime0} 2k^0
 2p^{\prime0} 2p^0}}  
 \varphi^{*}(k^\prime,p^\prime) \varphi(k,p) 
 \\ \times
 \text{Tr} \big\{(\vec{a}\cdot\boldsymbol{\sigma}) \mathcal{M}(k,p)
 (\vec{b}\cdot\boldsymbol{\sigma}^{T}) 
 \mathcal{M}^{\dagger}(k^\prime,p^\prime) \big\} 
 \\ \times
  \Delta_A(\vec{k}^\prime-\vec{k}) 
 \Delta_B(\vec{p}^\prime-\vec{p}),
\end{multline}
where Eq.~(\ref{amp}) implies
\begin{widetext}
\begin{multline}
 \text{Tr} \big\{ (\vec{a}\cdot\boldsymbol{\sigma})
 \mathcal{M}(k,p) (\vec{b}\cdot\boldsymbol{\sigma}^{T})
 \mathcal{M}^{\dagger}(k^\prime,p^\prime) \big\} =
 -\Big( 2 m^2 \sqrt{(m+p^0) (m+k^0) 
 (m+p^{\prime0}) (m+k^{\prime0})}\Big)^{-1} 
 \\ \times
 \Big\{ [\vec{a}\cdot(\vec{k}\times\vec{p})]
 [\vec{b}\cdot(\vec{k}^\prime\times\vec{p}^\prime)]
 + [\vec{a}\cdot(\vec{k}^\prime\times\vec{p}^\prime)]
 [\vec{b}\cdot(\vec{k}\times\vec{p})]
 - (\vec{a}\cdot\vec{b}) 
 [(\vec{k}\times\vec{p})\cdot(\vec{k}^\prime\times\vec{p}^\prime)]
 \\
 + (\vec{a}\cdot\vec{b})
 [(m+k^0)(m+p^0) - \vec{k}\cdot\vec{p}]
 [(m+k^{\prime0})(m+p^{\prime0}) 
 - \vec{k}^\prime\cdot\vec{p}^\prime]
 \\ 
 - (\vec{a}\times\vec{b}) \cdot
 \Big[ (\vec{k}^\prime\times\vec{p}^\prime)
 [(m+k^0)(m+p^0) - \vec{k}\cdot\vec{p}]
 +(\vec{k}\times\vec{p})
 [(m+k^{\prime0})(m+p^{\prime0})
 - \vec{k}^\prime\cdot\vec{p}^\prime]\Big]\Big\}.
 \label{slad_a_b_general}
\end{multline}
\end{widetext}
The denominator of the right hand side of Eq.~(\ref{eq:korelacje_omega})
takes the form 
\begin{multline}
 \label{eq:korelacja_mianownik}
 \bra{\varphi} \hat{\Pi}_A \otimes \hat{\Pi}_B \ket{\varphi} 
 \\ =
 \int \frac{d^3\vec{k}^\prime\,d^3 \vec{k}\, d^3 \vec{p}^\prime\,
  d^3 \vec{p}}{\sqrt{2k^{\prime0} 2k^0 2p^{\prime0} 2p^0}}
 \varphi^{*}(k^\prime,p^\prime) \varphi(k,p)
 \\ \times
 \text{Tr} \big\{\mathcal{M}(k,p) 
 \mathcal{M}(k^\prime,p^\prime)^{\dagger}\big\}
 \Delta_A(\vec{k}^\prime-\vec{k})\Delta_B(\vec{p}^\prime-\vec{p}),
\end{multline}
where
\begin{multline}
 \label{eq:trace_bez_sigm}
 \text{Tr} \big\{ \mathcal{M}(k,p) 
 \mathcal{M}^{\dagger}(k^\prime,p^\prime) \big\} 
 \\ =
 \Big( 
 2m^2 \sqrt{(m+p^0)(m+k^0)(m+p^{\prime0})(m+k^{\prime0})}
 \Big)^{-1}
 \\ \times
 \Big\{[(m+k^0)(m+p^0)-\vec{k}\cdot\vec{p}]
 \\ \times
 [(m+k^{\prime0})(m+p^{\prime0})-\vec{k}^\prime\cdot\vec{p}^\prime]
 + (\vec{k}\times\vec{p})
 \cdot(\vec{k}^\prime\times\vec{p}^\prime)\Big\}.
\end{multline}
We are now going to consider some special cases of
(\ref{eq:korelacje_omega}). From now on, we take the wave function
$\varphi(k,p)$ to be of the following form: 
\begin{equation}
 \label{fi}
 \varphi(k,p) = \varphi(k) \varphi(p).
\end{equation}
Adopting (\ref{fi}) does not imply that our state is separable in
momenta---entanglement is still present in the structure of
$\mathcal{M}(k,p)$, as given by Eq.~(\ref{amp}). With this assumption, it
is easily checked that (\ref{eq:korelacja_licznik}) and
(\ref{eq:korelacja_mianownik}) may be reduced to products of integrals
of three types: 
\begin{subequations}
\label{calki_gen}
\begin{multline}
 I_{1}^{A}[\varphi] =
 \int 
 \frac{d^3 \vec{k}^\prime\, d^3 \vec{k} \,
 (m+k^0)(m+k^{\prime0})}{
 \sqrt{k^0(m+k^0)}\sqrt{k^{\prime0}(m+k^{\prime0})}} 
 \\ \times
 \Delta_{A}(\vec{k}^\prime-\vec{k})
 \varphi^{*}(k^\prime) \varphi(k),
\end{multline}
\begin{multline}
 I_{2}^{A\,i}[\varphi] =
 \int 
 \frac{d^3 \vec{k}^\prime\, d^3 \vec{k} \,
 k^i (m+k^{\prime0})}{
 \sqrt{k^0(m+k^0)}\sqrt{k^{\prime0}(m+k^{\prime0})}} 
 \\ \times
 \Delta_{A}(\vec{k}^\prime-\vec{k})
 \varphi^{*}(k^\prime) \varphi(k),
\end{multline}
\begin{equation}
 I_{3}^{A\,ij}[\varphi] =
 \int 
 \frac{d^3 \vec{k}^\prime\, d^3 \vec{k} \,
 k^i k^{\prime j}\,
 \Delta_{A}(\vec{k}^\prime-\vec{k})
 \varphi^{*}(k^\prime) \varphi(k)}{
 \sqrt{k^0(m+k^0)}\sqrt{k^{\prime0}(m+k^{\prime0})}} 
\end{equation}
\end{subequations}
taken with appropriate coefficients. 
For Eq.~(\ref{eq:korelacja_licznik}) we have
\begin{widetext}
\begin{multline}
 \label{eq:korelacja_licznik_special}
 \bra{\varphi} \hat{\Pi}_A \big(\vec{a}\cdot\hat{\vec{S}}\big)
 \otimes \big(\vec{b}\cdot\hat{\vec{S}}\big) 
 \hat{\Pi}_B \ket{\varphi} 
 = \frac{-1}{2^5m^2} 
 \Big\{ (\vec{a}\cdot\vec{b}) \Big[
 I_{1}^{A} I_{1}^{B}
 -(\vec{I}_{2}^{A}\cdot\vec{I}_{2}^{B}) 
 -(\vec{I}_{2}^{A}\cdot\vec{I}_{2}^{B})^*
 +\text{Tr} \big[I_{3}^{A} I_{3}^{B*}\big]
 +\text{Tr} \big[I_{3}^{A} I_{3}^{B}\big]
 -\text{Tr} \big[I_{3}^{A}\big]
 \text{Tr} \big[I_{3}^{B}\big]
 \Big]
 \\
 -(\vec{a}\times\vec{b}) \cdot \Big[
 (\vec{I}_{2}^{A}\times\vec{I}_{2}^{B})
 + (\vec{I}_{2}^{A}\times\vec{I}_{2}^{B})^*
 \Big]
 + \vec{a}^T \Big[ 
 I_{3}^{A} (I_{3}^{B})^*
 - I_{3}^{B} (I_{3}^{A})^*
 + (I_{3}^{A})^* I_{3}^{B}
 - (I_{3}^{B})^* I_{3}^{A}
 \Big] \vec{b}
 \\ 
 + \varepsilon^{ijk} \varepsilon^{qrs} a^i b^q
 \Big[
 I_{3}^{A\,jr} I_{3}^{B\,ks}
 + (I_{3}^{A\,jr} I_{3}^{B\,ks})^*
 \Big] \Big\},
\end{multline}
\end{widetext}
while for
(\ref{eq:korelacja_mianownik}) we have: 
\begin{multline}
 \bra{\varphi}\hat{\Pi}_A \otimes \hat{\Pi}_B \ket{\varphi}
 = \frac{1}{2^3 m^2}
 \big\{
 I_{1}^{A} I_{1}^{B}
 -(\vec{I}_{2}^{A}\cdot\vec{I}_{2}^{B}) 
 -(\vec{I}_{2}^{A}\cdot\vec{I}_{2}^{B})^*
 \\ 
 +\text{Tr} \big[I_{3}^{A} I_{3}^{B*}\big]
 -\text{Tr} \big[I_{3}^{A} I_{3}^{B}\big]
 +\text{Tr} \big[I_{3}^{A}\big]
 \text{Tr} \big[I_{3}^{B}\big] \big\}.
 \label{eq:korelacja_mianownik_special}
\end{multline}


Now let us consider the simplest case of two particles with sharp
momenta. Therefore, we assume that
\begin{equation}
 \varphi(k,p) \to 
 2q_{a}^0 \delta^{3}(\vec{k}-\vec{q}_a)\,
 2q_{b}^0 \delta^{3}(\vec{p}-\vec{q}_b),
\end{equation}
where by $q_a$ and $q_b$ we have denoted the fixed four-momenta of
particles $a$ and $b$, respectively.
Thus our state takes the form
\begin{equation}
 \label{eg:stan_corelacje2}
 \ket{\varphi} \to
 \ket{\varphi_{q_a q_b}} =
 \mathcal{M}(q_a,q_b)_{\sigma\lambda} 
 \ket{q_a,\sigma}\otimes\ket{q_b,\lambda}.
\end{equation}
The numerator of the correlation function (\ref{eq:korelacje_omega}) reads:
\begin{multline}
 \label{eq:korelacja2}
 \bra{\varphi_{q_aq_b}} \big(\vec{a}\cdot\hat{\vec{S}}_A\big)
 \big(\vec{b}\cdot\hat{\vec{S}}_B\big) \ket{\varphi_{q_aq_b}}
 \\
 = \frac{q_a^0 q_b^0}{(2\pi)^6}
 \text{Tr} 
 \big\{(\vec{a}\cdot\boldsymbol{\sigma})
 \mathcal{M}(q_a,q_b) (\vec{b}\cdot\boldsymbol{\sigma}^{T})
 \mathcal{M}^{\dagger}(q_a,q_b)\big\}
 \\ \times
 \text{Vol}(A)\text{Vol}(B),
\end{multline}
while the denominator has the form
\begin{multline}
 \label{eq:korelacja_mianownik_ust_kier}
 \bra{\varphi_{q_aq_b}} \hat{\Pi}_A \otimes \hat{\Pi}_B 
 \ket{\varphi_{q_aq_b}} =
 \frac{4 q_a^0 q_b^0}{(2\pi)^6}
 \\ \times
 \text{Tr} 
 \big\{\mathcal{M}(q_a,q_b)\mathcal{M}^{\dagger}(q_a,q_b)\big\}
 \text{Vol}(A)\text{Vol}(B).
\end{multline}
By dividing the above formulas we receive
\begin{equation}
 \mathcal{C}_{\varphi_{q_aq_b}}^{AB}(\vec{a},\vec{b}) = 
 \frac{\text{Tr} 
 \big\{(\vec{a}\cdot\boldsymbol{\sigma})
 \mathcal{M}(q_a,q_b) (\vec{b}\cdot\boldsymbol{\sigma}^{T})
 \mathcal{M}^{\dagger}(q_a,q_b)\big\}}{\text{Tr} 
 \big\{\mathcal{M}(q_a,q_b)\mathcal{M}^{\dagger}(q_a,q_b)\big\}}
\end{equation}
and inserting the explicit form of the matrix $\mathcal{M}(q_a,q_b)$,
Eq.~(\ref{amp}), we finally get
\begin{multline}
 \mathcal{C}_{\varphi_{q_aq_b}}^{AB}(\vec{a},\vec{b}) = 
 -\vec{a}\cdot\vec{b} 
 + \frac{(\vec{q}_a\times\vec{q}_b)}{m^2+q_aq_b}\cdot
 \bigg((\vec{a}\times\vec{b}) \\
   +\frac{(\vec{a}\cdot\vec{q}_a)(\vec{b}\times\vec{q}_b) -
     (\vec{b}\cdot\vec{q}_b)
 (\vec{a}\times\vec{q}_a)}{(q_a^0+m)(q_b^0+m)}\bigg).
 \label{correl_sharp_mom}
\end{multline}
Comparing the correlation function (\ref{correl_sharp_mom}) with the
previous results for the correlation function without localization,
\cite{CR2005,CR2006}, we see that for sharp momentum states
localization inside detectors does not change the correlation
function.


Now let us consider more general situation in which only the
directions of particle momenta are fixed. 
Thus, let us denote directions of the momenta of the first and second
particle by $\vec{n}$ and $\vec{m}$, respectively.
We assume that the wave function has the following form:
\begin{multline}
 \varphi(k,p) \to
 \frac{\sqrt{k^0(m+k^0)}}{\vec{k}^2} f(|\vec{k}|) 
 \delta(\tfrac{\vec{k}}{|\vec{k}|}-\vec{n})
 \\ \times
 \frac{\sqrt{p^0(m+p^0)}}{\vec{p}^2} f(|\vec{p}|) 
 \delta(\tfrac{\vec{p}}{|\vec{p}|}-\vec{m}),
\end{multline}
where $\delta(\tfrac{\vec{k}}{|\vec{k}|}-\vec{n})$ is a Dirac delta
projecting on a fixed direction, i.e.
\begin{equation}
 \int d\Omega(\alpha,\beta) \delta(\vec{n}(\alpha,\beta)-\vec{n})
 g(\vec{n}(\alpha,\beta)) = g(\vec{n}),
\end{equation}
where $d\Omega(\alpha,\beta)$ is a differential solid angle.

In the considered case the correlation function
(\ref{eq:korelacje_omega}) can be expressed in terms of the following
integrals
\begin{subequations}
\begin{multline}
 I_{1}^{A,\vec{n}} = \int_{0}^{\infty} dt\, du\, 
 (m+\sqrt{m^2+t^2}) (m+\sqrt{m^2+u^2})
 \\ \times
 \Delta_A((t-u)\vec{n}) f^*(t) f(u),
\end{multline}
\begin{multline}
 I_{2}^{A,\vec{n}} = \int_{0}^{\infty} dt\, du\, 
 u (m+\sqrt{m^2+t^2})
 \\ \times
 \Delta_A((t-u)\vec{n}) f^*(t) f(u),
\end{multline}
\begin{equation}
 I_{3}^{A,\vec{n}} = \int_{0}^{\infty} dt\, du\, 
 tu \Delta_A((t-u)\vec{n}) f^*(t) f(u).
\end{equation}
\end{subequations}
Indeed, in this case integrals (\ref{calki_gen}) are equal to
\begin{subequations}
\begin{align}
& I_{1}^{A} = I_{1}^{A,\vec{n}}, 
&&I_{1}^{B} = I_{1}^{B,\vec{m}}, \\
&I_{2}^{A\,i} = n^i I_{2}^{A,\vec{n}}, 
&&I_{2}^{B\,i} = m^i I_{2}^{B,\vec{m}}, \\
&I_{3}^{A,ij} = n^i n^j I_{3}^{A,\vec{n}}, 
&&I_{3}^{B\,ij} = m^i m^j I_{3}^{B,\vec{m}}. 
\end{align}
\end{subequations}
Therefore, for the correlation function, with the help of
Eqs.~(\ref{eq:korelacja_licznik_special},\ref{eq:korelacja_mianownik_special})
and (\ref{eq:korelacje_omega}), we receive
\begin{widetext}
\begin{multline}
 \mathcal{C}_{\vec{n},\vec{m}}^{AB}(\vec{a},\vec{b})
 = \Big\{
 -(\vec{a}\cdot\vec{b}) I_{1}^{A,\vec{n}} I_{1}^{B,\vec{m}}
  +\big[
 (\vec{a}\cdot\vec{b}) (\vec{n}\times\vec{m})^2
 - (\vec{a}\cdot\vec{b}) (\vec{n}\cdot\vec{m})^2
  - 2 [\vec{a}\cdot(\vec{n}\times\vec{m})]
 [\vec{b}\cdot(\vec{n}\times\vec{m})]
 \\
 - 2 (\vec{a}\times\vec{b})\cdot(\vec{n}\times\vec{m})
 (\vec{n}\cdot\vec{m})
 \big] 
 I_{3}^{A,\vec{n}} I_{3}^{B,\vec{m}}
 +\big[  
 (\vec{a}\cdot\vec{b})(\vec{n}\cdot\vec{m})
 + (\vec{a}\times\vec{b})\cdot(\vec{n}\times\vec{m})
 \big] 
 \big[
 I_{2}^{A,\vec{n}} I_{2}^{B,\vec{m}}
 + (I_{2}^{A,\vec{n}} I_{2}^{B,\vec{m}})^*
 \big]
 \Big\}
 \\ \times
 \Big\{
 I_{1}^{A,\vec{n}} I_{1}^{B,\vec{m}}
 + I_{3}^{A,\vec{n}} I_{3}^{B,\vec{m}}
 - (\vec{n}\cdot\vec{m})
 \big[
 I_{2}^{A,\vec{n}} I_{2}^{B,\vec{m}}
 + (I_{2}^{A,\vec{n}} I_{2}^{B,\vec{m}})^*
 \big]
 \Big\}^{-1}.
\end{multline}
\end{widetext}
It is worth to stress that when $\vec{m}=-\vec{n}$, the above
correlation function is equal to
\begin{equation}
 \mathcal{C}_{\vec{n},-\vec{n}}^{AB}(\vec{a},\vec{b}) =
 -\vec{a}\cdot\vec{b}.
\end{equation}
Thus, for particles propagating in opposite directions
loczlization inside detectors does not change the correlation
function.

In conclusion, we have derived the correlation function in an
arbitrary scalar state of two fermions assuming that spin projection  
measurement is associated with the localization of the particles
inside detectors
[Eqs.~(\ref{eq:korelacje_omega}-\ref{eq:trace_bez_sigm})].
We have also shown that in some situations the
localization does not change the correlation function.
It is the case when the wave function fulfils the condition
(\ref{fi}) and (i) particles have sharp momenta or (ii) directions of
particles momenta are fixed and correlation function is measured in
the center-of-mass frame.


\begin{acknowledgments}
This work has been supported by the University of Lodz and by the Polish
Ministry of Science and Higher Education under the contract
No.~N~N202~103738. 
\end{acknowledgments}


\begin{thebibliography}{18}%
\makeatletter
\providecommand \@ifxundefined [1]{%
 \@ifx{#1\undefined}
}%
\providecommand \@ifnum [1]{%
 \ifnum #1\expandafter \@firstoftwo
 \else \expandafter \@secondoftwo
 \fi
}%
\providecommand \@ifx [1]{%
 \ifx #1\expandafter \@firstoftwo
 \else \expandafter \@secondoftwo
 \fi
}%
\providecommand \natexlab [1]{#1}%
\providecommand \enquote  [1]{``#1''}%
\providecommand \bibnamefont  [1]{#1}%
\providecommand \bibfnamefont [1]{#1}%
\providecommand \citenamefont [1]{#1}%
\providecommand \href@noop [0]{\@secondoftwo}%
\providecommand \href [0]{\begingroup \@sanitize@url \@href}%
\providecommand \@href[1]{\@@startlink{#1}\@@href}%
\providecommand \@@href[1]{\endgroup#1\@@endlink}%
\providecommand \@sanitize@url [0]{\catcode `\\12\catcode `\$12\catcode
  `\&12\catcode `\#12\catcode `\^12\catcode `\_12\catcode `\%12\relax}%
\providecommand \@@startlink[1]{}%
\providecommand \@@endlink[0]{}%
\providecommand \url  [0]{\begingroup\@sanitize@url \@url }%
\providecommand \@url [1]{\endgroup\@href {#1}{\urlprefix }}%
\providecommand \urlprefix  [0]{URL }%
\providecommand \Eprint [0]{\href }%
\providecommand \doibase [0]{http://dx.doi.org/}%
\providecommand \selectlanguage [0]{\@gobble}%
\providecommand \bibinfo  [0]{\@secondoftwo}%
\providecommand \bibfield  [0]{\@secondoftwo}%
\providecommand \translation [1]{[#1]}%
\providecommand \BibitemOpen [0]{}%
\providecommand \bibitemStop [0]{}%
\providecommand \bibitemNoStop [0]{.\EOS\space}%
\providecommand \EOS [0]{\spacefactor3000\relax}%
\providecommand \BibitemShut  [1]{\csname bibitem#1\endcsname}%
\let\auto@bib@innerbib\@empty
\bibitem [{\citenamefont {Caban}\ and\ \citenamefont
  {Rembieli\'nski}(2006)}]{CR2006}%
  \BibitemOpen
  \bibfield  {author} {\bibinfo {author} {\bibfnamefont {P.}~\bibnamefont
  {Caban}}\ and\ \bibinfo {author} {\bibfnamefont {J.}~\bibnamefont
  {Rembieli\'nski}},\ }\href@noop {} {\bibfield  {journal} {\bibinfo  {journal}
  {Phys. Rev. A}\ }\textbf {\bibinfo {volume} {74}},\ \bibinfo {pages} {042103}
  (\bibinfo {year} {2006})}\BibitemShut {NoStop}%
\bibitem [{\citenamefont {Friis}\ \emph {et~al.}(2010)\citenamefont {Friis},
  \citenamefont {Bertlmann}, \citenamefont {Huber},\ and\ \citenamefont
  {Hiesmayr}}]{FBHH2010}%
  \BibitemOpen
  \bibfield  {author} {\bibinfo {author} {\bibfnamefont {N.}~\bibnamefont
  {Friis}}, \bibinfo {author} {\bibfnamefont {R.~A.}\ \bibnamefont
  {Bertlmann}}, \bibinfo {author} {\bibfnamefont {M.}~\bibnamefont {Huber}}, \
  and\ \bibinfo {author} {\bibfnamefont {B.~C.}\ \bibnamefont {Hiesmayr}},\
  }\href@noop {} {\bibfield  {journal} {\bibinfo  {journal} {Phys. Rev. A}\
  }\textbf {\bibinfo {volume} {81}},\ \bibinfo {pages} {042114} (\bibinfo
  {year} {2010})}\BibitemShut {NoStop}%
\bibitem [{\citenamefont {Czachor}(2010)}]{Czachor2010}%
  \BibitemOpen
  \bibfield  {author} {\bibinfo {author} {\bibfnamefont {M.}~\bibnamefont
  {Czachor}},\ }\href@noop {} {\bibfield  {journal} {\bibinfo  {journal}
  {Quantum Inf Process}\ }\textbf {\bibinfo {volume} {9}},\ \bibinfo {pages}
  {171} (\bibinfo {year} {2010})}\BibitemShut {NoStop}%
\bibitem [{\citenamefont {Saldanha}\ and\ \citenamefont
  {Vedral}(2012{\natexlab{a}})}]{SV2011}%
  \BibitemOpen
  \bibfield  {author} {\bibinfo {author} {\bibfnamefont {P.~L.}\ \bibnamefont
  {Saldanha}}\ and\ \bibinfo {author} {\bibfnamefont {V.}~\bibnamefont
  {Vedral}},\ }\href@noop {} {\bibfield  {journal} {\bibinfo  {journal} {Phys.
  Rev. A}\ }\textbf {\bibinfo {volume} {85}},\ \bibinfo {pages} {062101}
  (\bibinfo {year} {2012}{\natexlab{a}})}\BibitemShut {NoStop}%
\bibitem [{\citenamefont {Debarba}\ and\ \citenamefont
  {Vianna}(2012)}]{DV2012}%
  \BibitemOpen
  \bibfield  {author} {\bibinfo {author} {\bibfnamefont {T.}~\bibnamefont
  {Debarba}}\ and\ \bibinfo {author} {\bibfnamefont {R.~O.}\ \bibnamefont
  {Vianna}},\ }\href@noop {} {\bibfield  {journal} {\bibinfo  {journal} {Int.
  J. Quant. Inf.}\ }\textbf {\bibinfo {volume} {10}},\ \bibinfo {pages}
  {1230003} (\bibinfo {year} {2012})}\BibitemShut {NoStop}%
\bibitem [{\citenamefont {Caban}\ \emph {et~al.}(2009)\citenamefont {Caban},
  \citenamefont {Rembieli\'nski},\ and\ \citenamefont
  {W{\l}odarczyk}}]{CRW2009}%
  \BibitemOpen
  \bibfield  {author} {\bibinfo {author} {\bibfnamefont {P.}~\bibnamefont
  {Caban}}, \bibinfo {author} {\bibfnamefont {J.}~\bibnamefont
  {Rembieli\'nski}}, \ and\ \bibinfo {author} {\bibfnamefont {M.}~\bibnamefont
  {W{\l}odarczyk}},\ }\href@noop {} {\bibfield  {journal} {\bibinfo  {journal}
  {Phys. Rev. A}\ }\textbf {\bibinfo {volume} {79}},\ \bibinfo {pages} {014102}
  (\bibinfo {year} {2009})}\BibitemShut {NoStop}%
\bibitem [{\citenamefont {Caban}(2007)}]{Caban2007_photons}%
  \BibitemOpen
  \bibfield  {author} {\bibinfo {author} {\bibfnamefont {P.}~\bibnamefont
  {Caban}},\ }\href@noop {} {\bibfield  {journal} {\bibinfo  {journal} {Phys.
  Rev. A}\ }\textbf {\bibinfo {volume} {76}},\ \bibinfo {pages} {052102}
  (\bibinfo {year} {2007})}\BibitemShut {NoStop}%
\bibitem [{\citenamefont {Czachor}(1997)}]{Czachor1997_1}%
  \BibitemOpen
  \bibfield  {author} {\bibinfo {author} {\bibfnamefont {M.}~\bibnamefont
  {Czachor}},\ }\href@noop {} {\bibfield  {journal} {\bibinfo  {journal} {Phys.
  Rev. A}\ }\textbf {\bibinfo {volume} {55}},\ \bibinfo {pages} {72} (\bibinfo
  {year} {1997})}\BibitemShut {NoStop}%
\bibitem [{\citenamefont {Terno}(2003)}]{Terno2003}%
  \BibitemOpen
  \bibfield  {author} {\bibinfo {author} {\bibfnamefont {D.~R.}\ \bibnamefont
  {Terno}},\ }\href@noop {} {\bibfield  {journal} {\bibinfo  {journal} {Phys.
  Rev. A}\ }\textbf {\bibinfo {volume} {67}},\ \bibinfo {pages} {014102}
  (\bibinfo {year} {2003})}\BibitemShut {NoStop}%
\bibitem [{\citenamefont {Caban}\ and\ \citenamefont
  {Rembieli\'nski}(2005)}]{CR2005}%
  \BibitemOpen
  \bibfield  {author} {\bibinfo {author} {\bibfnamefont {P.}~\bibnamefont
  {Caban}}\ and\ \bibinfo {author} {\bibfnamefont {J.}~\bibnamefont
  {Rembieli\'nski}},\ }\href@noop {} {\bibfield  {journal} {\bibinfo  {journal}
  {Phys. Rev. A}\ }\textbf {\bibinfo {volume} {72}},\ \bibinfo {pages} {012103}
  (\bibinfo {year} {2005})}\BibitemShut {NoStop}%
\bibitem [{\citenamefont {Saldanha}\ and\ \citenamefont
  {Vedral}(2012{\natexlab{b}})}]{SV2012}%
  \BibitemOpen
  \bibfield  {author} {\bibinfo {author} {\bibfnamefont {P.~L.}\ \bibnamefont
  {Saldanha}}\ and\ \bibinfo {author} {\bibfnamefont {V.}~\bibnamefont
  {Vedral}},\ }\href@noop {} {\bibfield  {journal} {\bibinfo  {journal} {New J.
  Phys.}\ }\textbf {\bibinfo {volume} {14}},\ \bibinfo {pages} {023041}
  (\bibinfo {year} {2012}{\natexlab{b}})}\BibitemShut {NoStop}%
\bibitem [{\citenamefont {Palmer}\ \emph {et~al.}(2013)\citenamefont {Palmer},
  \citenamefont {Takahashi},\ and\ \citenamefont {Westman}}]{PTW2013_spin_WKB}%
  \BibitemOpen
  \bibfield  {author} {\bibinfo {author} {\bibfnamefont {M.~C.}\ \bibnamefont
  {Palmer}}, \bibinfo {author} {\bibfnamefont {M.}~\bibnamefont {Takahashi}}, \
  and\ \bibinfo {author} {\bibfnamefont {H.~F.}\ \bibnamefont {Westman}},\
  }\href@noop {} {\bibfield  {journal} {\bibinfo  {journal} {Annals of Phys.}\
  }\textbf {\bibinfo {volume} {336}},\ \bibinfo {pages} {505} (\bibinfo {year}
  {2013})}\BibitemShut {NoStop}%
\bibitem [{\citenamefont {Bacry}(1988)}]{cab_Bacry1988}%
  \BibitemOpen
  \bibfield  {author} {\bibinfo {author} {\bibfnamefont {H.}~\bibnamefont
  {Bacry}},\ }\href@noop {} {\emph {\bibinfo {title} {Localizability and Space
  in Quantum Physics}}},\ Lecture Notes in Physics Vol.\ 308\ (\bibinfo
  {publisher} {Springer--Verlag},\ \bibinfo {address} {Berlin, Heidelberg},\
  \bibinfo {year} {1988})\BibitemShut {NoStop}%
\bibitem [{\citenamefont {Caban}\ \emph
  {et~al.}(2013{\natexlab{a}})\citenamefont {Caban}, \citenamefont
  {Rembieli\'nski},\ and\ \citenamefont
  {W{\l}odarczyk}}]{CRW_2013_Dirac_spin_PRA}%
  \BibitemOpen
  \bibfield  {author} {\bibinfo {author} {\bibfnamefont {P.}~\bibnamefont
  {Caban}}, \bibinfo {author} {\bibfnamefont {J.}~\bibnamefont
  {Rembieli\'nski}}, \ and\ \bibinfo {author} {\bibfnamefont {M.}~\bibnamefont
  {W{\l}odarczyk}},\ }\href@noop {} {\bibfield  {journal} {\bibinfo  {journal}
  {Phys. Rev. A}\ }\textbf {\bibinfo {volume} {88}},\ \bibinfo {pages} {022119}
  (\bibinfo {year} {2013}{\natexlab{a}})}\BibitemShut {NoStop}%
\bibitem [{\citenamefont {Newton}\ and\ \citenamefont
  {Wigner}(1949)}]{cab_NW1949}%
  \BibitemOpen
  \bibfield  {author} {\bibinfo {author} {\bibfnamefont {T.~D.}\ \bibnamefont
  {Newton}}\ and\ \bibinfo {author} {\bibfnamefont {E.~P.}\ \bibnamefont
  {Wigner}},\ }\href@noop {} {\bibfield  {journal} {\bibinfo  {journal} {Rev.
  Mod. Phys}\ }\textbf {\bibinfo {volume} {21}},\ \bibinfo {pages} {400}
  (\bibinfo {year} {1949})}\BibitemShut {NoStop}%
\bibitem [{\citenamefont {Foldy}\ and\ \citenamefont
  {Wouthuysen}(1950)}]{FW1950}%
  \BibitemOpen
  \bibfield  {author} {\bibinfo {author} {\bibfnamefont {L.~L.}\ \bibnamefont
  {Foldy}}\ and\ \bibinfo {author} {\bibfnamefont {S.~A.}\ \bibnamefont
  {Wouthuysen}},\ }\href@noop {} {\bibfield  {journal} {\bibinfo  {journal}
  {Phys. Rev.}\ }\textbf {\bibinfo {volume} {78}},\ \bibinfo {pages} {29}
  (\bibinfo {year} {1950})}\BibitemShut {NoStop}%
\bibitem [{\citenamefont {Bogolubov}\ \emph {et~al.}(1975)\citenamefont
  {Bogolubov}, \citenamefont {Logunov},\ and\ \citenamefont
  {Todorov}}]{cab_BLT1969}%
  \BibitemOpen
  \bibfield  {author} {\bibinfo {author} {\bibfnamefont {N.~N.}\ \bibnamefont
  {Bogolubov}}, \bibinfo {author} {\bibfnamefont {A.~A.}\ \bibnamefont
  {Logunov}}, \ and\ \bibinfo {author} {\bibfnamefont {I.~T.}\ \bibnamefont
  {Todorov}},\ }\href@noop {} {\emph {\bibinfo {title} {Introduction to
  Axiomatic Quantum Field Theory}}}\ (\bibinfo  {publisher} {W. A. Benjamin},\
  \bibinfo {address} {Reading, Mass.},\ \bibinfo {year} {1975})\BibitemShut
  {NoStop}%
\bibitem [{\citenamefont {Caban}\ \emph
  {et~al.}(2013{\natexlab{b}})\citenamefont {Caban}, \citenamefont
  {Rembieli\'nski},\ and\ \citenamefont {W{\l}odarczyk}}]{CRW_2013_Dirac_spin}%
  \BibitemOpen
  \bibfield  {author} {\bibinfo {author} {\bibfnamefont {P.}~\bibnamefont
  {Caban}}, \bibinfo {author} {\bibfnamefont {J.}~\bibnamefont
  {Rembieli\'nski}}, \ and\ \bibinfo {author} {\bibfnamefont {M.}~\bibnamefont
  {W{\l}odarczyk}},\ }\href@noop {} {\bibfield  {journal} {\bibinfo  {journal}
  {Annals of Phys.}\ }\textbf {\bibinfo {volume} {330}},\ \bibinfo {pages}
  {263} (\bibinfo {year} {2013}{\natexlab{b}})}\BibitemShut {NoStop}%
\end{thebibliography}

%

\end{document}